\documentclass{ifacconf}

\usepackage{graphicx}
\usepackage{natbib}
\usepackage{url}
\usepackage[dvipsnames]{xcolor}
\usepackage{subcaption}
\usepackage{booktabs}
\usepackage{tabularx}
\usepackage{amsmath,amssymb}
\usepackage{array}
\usepackage{rotating}
\usepackage{paralist}
\usepackage[inline]{enumitem}

\newcommand{\PPFD}{\mathrm{PPFD}}
\newcommand{\GHI}{\mathrm{GHI}}
\newcommand{\DLI}{\mathrm{DLI}}
\begin{document}
\begin{frontmatter}

\title{A Comparative Study of LED Lighting Control Strategies for Energy-Efficient Greenhouse Operations\thanksref{footnoteinfo}}

\thanks[footnoteinfo]{This work was supported by the USDA-NIFA Farm of the Future Program, under Award No. 2023-68017-39402.
}

\author[First]{Sajad Ahmadi}
\author[Second]{Eric J. Stallknecht}
\author[First]{Javad Mohammadpour Velni}

\address[First]{Dept. of Mechanical Engineering, Clemson University, Clemson, SC 29634, USA (e-mail: \{sahmadi, javadm\}@clemson.edu).}
\address[Second]{School of Plant and Environmental Sciences, Virginia Tech, 1444 Diamond Springs Road, Virginia Beach, VA 23455, USA (e-mail: ericjs3@vt.edu).}

\begin{abstract}
Greenhouse supplemental lighting is essential for meeting crop Daily Light Integral (DLI) requirement. However, conventional always-on and rule-based strategies are inherently energy-inefficient since they ignore future sunlight availability. This paper reports the results of deploying various supplemental lighting control strategies, namely, always-on, always-off, heuristic, optimal prediction-based, and two peak-shift variants, simultaneously in a single greenhouse at the Hampton Roads AREC of Virginia Tech, to ensure that every treatment experiences identical conditions. 
The prediction-based and peak-shift strategies optimize LED schedules using a Markov chain-based solar forecast. Experimental results over two replicates 
show that the prediction-based strategy meets the desired DLI target at 65\% cost reduction relative to the always-on baseline while achieving the highest dry-mass yield among the proposed adaptive strategies. The heuristic strategy incurs only a 6\% cost premium over prediction-based control, offering a competitive forecast-free alternative. The peak-shift variants achieve the lowest active-strategy costs by shifting LED demand into off-peak tariff windows. A yield-per-dollar efficiency metric is finally introduced to jointly capture energy cost and crop output showing that the prediction-based strategy ranks highest at 18.976~g/\$, more than double the always-on value of 9.391~g/\$. These findings can provide greenhouse operators with an evidence-based framework for selecting a lighting control approach that balances energy cost, crop quality, and system complexity.
\end{abstract}


\end{frontmatter}

\section{Introduction}
\label{sec:intro}

Light is one of the most critical environmental factors affecting plant growth, morphology, yield, and product quality in controlled-environment agriculture and greenhouse production. In many greenhouse systems, especially during winter months or cloudy periods, natural solar radiation is not sufficient to satisfy light requirements. As a result, supplemental lighting is used to maintain crop productivity, and shorten production cycles in plant development~\citep{Morrow2008,Faust2018,Singh2015}. Meanwhile, the resulting electricity cost constitutes a dominant share of the operating expenses, making the scheduling of lighting an immediately practical control problem.

The shift from high-pressure sodium fixtures to dimmable LED arrays has changed what is possible. LEDs offer high photon efficacy, millisecond-scale dimming, long life, and spectral tunability, and these properties reframe supplemental light from a fixed input to an actively controlled decision variable~\citep{Morrow2008,Singh2015}. Two quantities anchor any modern formulation: the instantaneous photosynthetic photon flux density (PPFD) at the canopy and the integrated Daily Light Integral (DLI). Well-designed controllers should deliver a target DLI rather than a fixed PPFD waveform, exploiting free sunlight whenever it is available and concentrating paid-for photons in the cheapest and most useful windows. A large body of work has pursued this goal via threshold and duty-cycle control~\citep{vanIersel2017}, photochemistry-based formulations~\citep{Weaver2019}, and optimization-based schemes coupling sunlight prediction with time-of-use (TOU) pricing~\citep{Mosharafian2021,Afzali2021}.

Three practical obstacles still block adoption. First, published studies evaluate different crops under different weather in different greenhouses, so cross-study comparisons conflate controller performance with experimental context. Second, the trade-offs among yield, energy cost, peak demand, and operational complexity are rarely quantified within a single consistent framework. Third, the optimization methods that perform best in simulation often require forecasting infrastructure whose marginal benefit over a simple rule is not well established empirically. This paper overcomes all three obstacles by deploying six lighting control strategies simultaneously in one greenhouse, on a shared Internet of Things (IoT) platform, with identical sensing and actuation hardware:
\begin{enumerate}
    \item \textbf{Always-on:} LEDs at fixed maximum intensity throughout the photoperiod.
    \item \textbf{Always-off:} no supplemental lighting (sunlight-only control).
    \item \textbf{Prediction-based:} an energy-minimizing linear program (LP) driven by a context-conditioned Markov-chain solar forecast.
    \item \textbf{Heuristic:} a forecast-free rule that drives LEDs to the catch-up rate needed to meet the DLI target in the remaining photoperiod.
    \item \textbf{Peak-shift start at 4:30 AM:} similar to prediction based with an added peak-power penalty, planning from 04:30.
    \item \textbf{Peak-shift start at 11:00 AM:} starts at 11:00 and minimizes peak LED power without TOU-cost weighting.
\end{enumerate}
Since every treatment sees identical weather and canopy environment, the only variable across treatments is the control law itself.\textbf{ The main contributions of this work are: \begin{enumerate*}[label=(\roman*)]
    \item A unified IoT platform that supports side-by-side deployment of multiple lighting strategies in a single greenhouse;
    \item A controlled experimental comparison of six strategies across energy cost, DLI delivery, and plant growth;
    \item A yield-per-dollar efficiency metric that captures the economic return of each control strategy.
\end{enumerate*}}

The remainder of this paper is organized as follows. Section~\ref{sec:system} describes the shared platform; Section~\ref{sec:greenhouse} describes the greenhouse layout; Section~\ref{sec:strategies} formalizes the six strategies within a single optimization framework; Section~\ref{sec:results} presents the comparative results, and finally, Section~\ref{sec:conclusion} concludes the paper.

\section{Experimental Platform}
\subsection{System Architecture}
\label{sec:system}
All six strategies share one hardware/software platform
with three layers (sensing, control,
and actuation); only the control layer differs across treatments.
Inter-component communication uses the Message Queuing Telemetry Transport (MQTT) protocol~\citep{MQTT2019},
whose lightweight publish-subscribe semantics suit resource-constrained
IoT devices. A dual-broker topology is adopted for robustness: a
\emph{cloud broker} handles device discovery and registration, while a
\emph{local broker} carries all real-time traffic (PPFD measurements,
forecasts, and LED commands) inside the greenhouse, keeping latency low
and preserving operation during internet outages.

Four Apogee SQ-500-SS full-spectrum quantum sensors (self-powered,
0--40~mV output over 0--4000~$\mu$mol\,m$^{-2}$\,s$^{-1}$) are deployed
along Block~B (Fig.~\ref{fig:greenhouse_layout}); their voltages are
converted to PPFD using the manufacturer-provided factory calibration
\begin{equation}
\PPFD_{\text{sens}}(t) \;=\; k_s \cdot V_{\text{sens}}(t),
\label{eq:sensor_cal}
\end{equation}
where $k_s = 100~\mu\text{mol}\,\text{m}^{-2}\,\text{s}^{-1}/\text{mV}$ is the sensor calibration factor. This value was not fitted using the treatment-response data; it was fixed before the lighting-control experiment and used unchanged for both replicates. The four converted PPFD signals were averaged into one representative indoor PPFD used as the feedback signal by all controllers.

LED intensity is set by pulse-width modulation (PWM), with an approximately linear duty-to-photon relationship
\begin{equation}
\PPFD_{\text{LED}}(t) \;=\; \gamma \cdot u_{\text{PWM}}(t),
\label{eq:pwm_to_ppfd}
\end{equation}
where $u_{\text{PWM}}(t) \in [0,1]$ and $\gamma = 261.29\ \mu$mol\,m$^{-2}$\,s$^{-1}$ is the calibrated full-duty LED gain. The value of $\gamma$ was obtained before the treatment runs by commanding the LED fixtures at fixed PWM levels under no-sunlight conditions, measuring the resulting canopy-level PPFD, and fitting the slope of LED PPFD versus PWM duty cycle. This calibration was completed separately from the crop-growth validation experiment; $\gamma$ was then held fixed for all strategies and both replicates. The driver input is therefore $\text{PWM}\,[\%] = (100/\gamma)\,\PPFD_{\text{LED}}$. All controllers refresh their LED command every $\Delta t = 300$\,s.


\subsection{Greenhouse Layout}
\label{sec:greenhouse}

\begin{figure}[!t]
    \centering
    \includegraphics[width=\columnwidth]{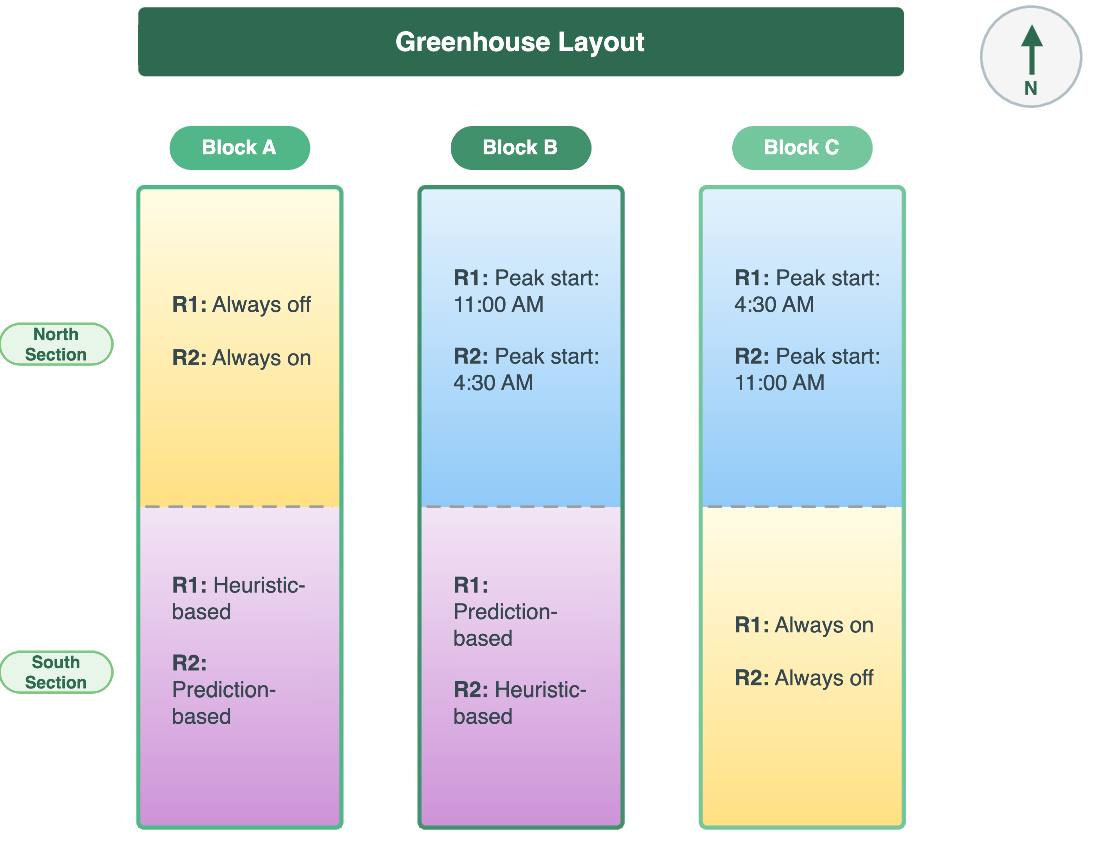}
    \caption{Top-view diagram of the greenhouse layout showing benching zones and strategy assignments. R1 and R2 denote replicate~1 and replicate~2; the two replicates are placed in diagonally opposite zones to mitigate shading and position effects.}
    \label{fig:greenhouse_layout}
\end{figure}

\begin{figure}[!t]
    \centering
    \begin{subfigure}[b]{0.8\columnwidth}
        \includegraphics[width=\linewidth]{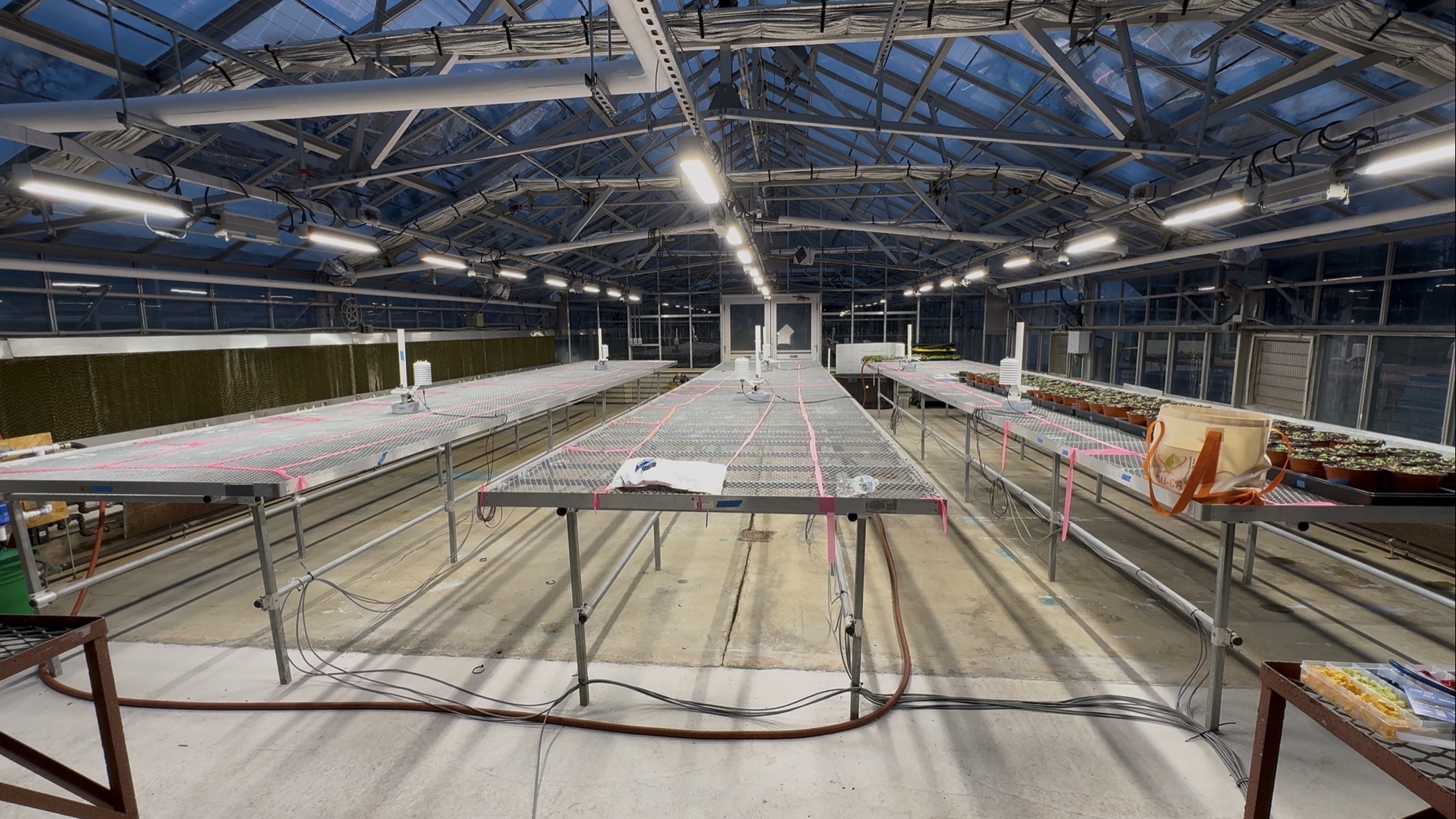}
        \caption{Interior view showing benching tables and LED mounting infrastructure from the south side.}
        \label{fig:greenhouse_interior}
    \end{subfigure}
    \hfill
    \begin{subfigure}[b]{0.8\columnwidth}
        \includegraphics[width=\linewidth]{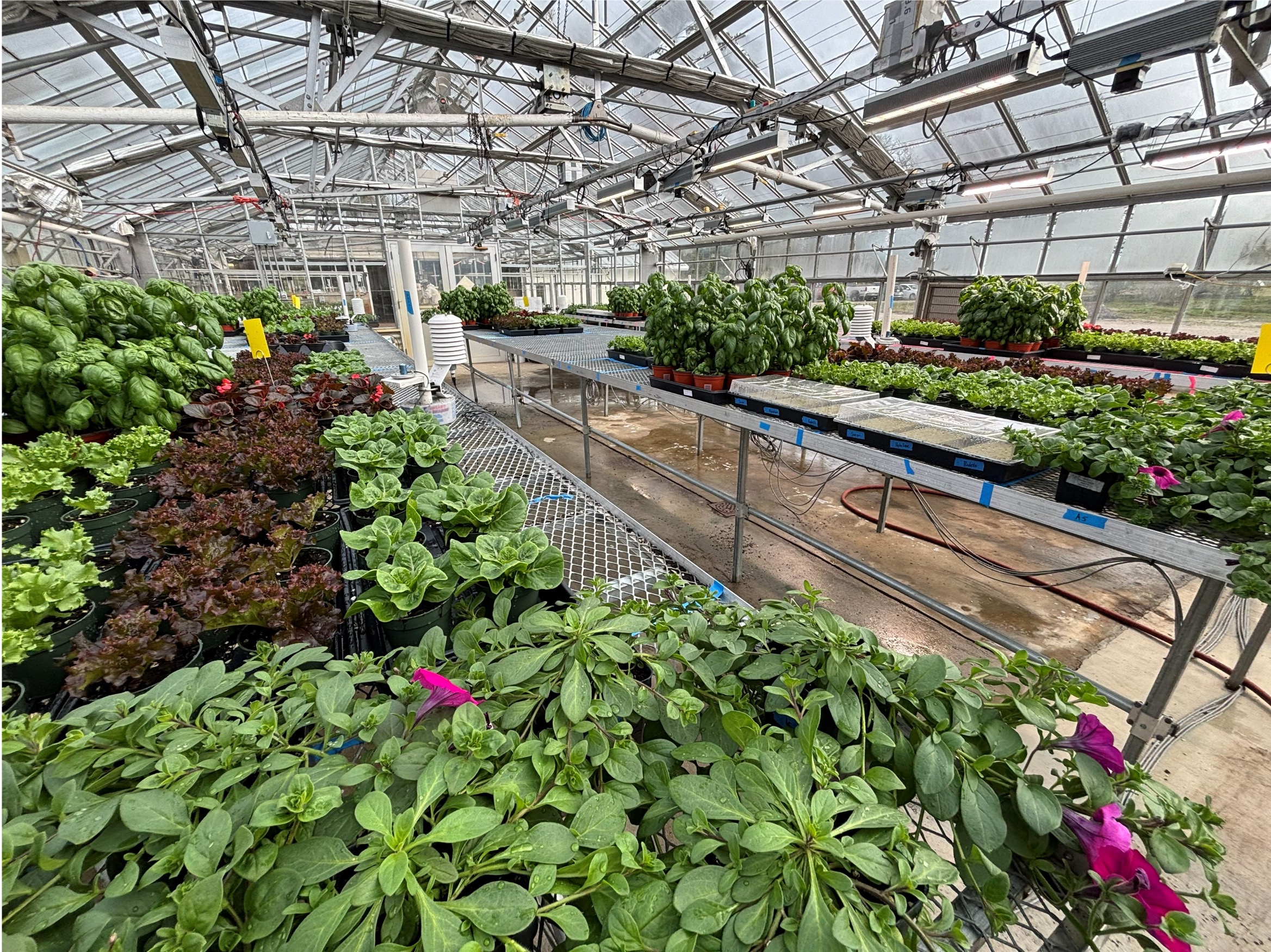}
        \caption{Interior of greenhouse one day before harvest.}
        \label{fig:greenhouse_preharvest}
    \end{subfigure}
    \caption{Interior views of the greenhouse.}
    \label{fig:greenhouse_views}
\end{figure}

The experiment was conducted at the Hampton Roads Agricultural Research and Extension Center (AREC) of Virginia Tech, Virginia Beach, VA\footnote{\url{https://www.arec.vaes.vt.edu/arec/hampton-roads.html}}. The experimental greenhouse is divided into six benching zones, each equipped with an independent LED array and sharing four common PPFD sensors. This modular layout enables simultaneous testing of different control strategies under identical ambient conditions. To mitigate potential shading and orientation effects, each strategy is replicated twice, with the two replicates placed in diagonally opposite zones (see Fig.~\ref{fig:greenhouse_layout} and Fig. \ref{fig:greenhouse_views}).

The crop used in the experiment was lettuce (\emph{Lactuca sativa} L.). Three cultivars were evaluated: Hydrolique, Rabello, and Sungari. The reported dry mass refers to the mean harvested above-ground shoot dry mass per plant after drying.

\section{Control Strategies and Experimental Design}
\label{sec:strategies}

This section first describes the sunlight forecasting model that is shared by the prediction-based and peak-shift strategies, and then formalizes the six control strategies.

\subsection{Markov Chain for PPFD Prediction}
\label{subsec:markov_ppfd}

Five-minute solar resource data were obtained from the NREL National Solar Radiation Database (NSRDB)~\citep{NSRDB2018} for the study location and arranged chronologically using local timestamps for the years 2018-2024. Although several meteorological variables were downloaded, the Markov-chain construction uses only the global horizontal irradiance (GHI) and the solar zenith angle. The outdoor PPFD is estimated from GHI as
\begin{equation}
\PPFD_t = \beta \, \GHI_t,
\label{eq:ghi_to_ppfd}
\end{equation}
with
\begin{equation}
\beta = 2.02 \times 0.662 \;=\; 1.337.
\label{eq:beta}
\end{equation}
The factor $2.02$ [$\mu$mol\,J$^{-1}$] converts GHI (in W/m\textsuperscript{2}) to outdoor PPFD (in $\mu$mol\,m$^{-2}$\,s$^{-1}$). Because the NSRDB values represent outdoor conditions, while the relevant quantity for plant growth is the indoor PPFD inside the greenhouse, a second factor of $0.662$ is included to account for light attenuation by the greenhouse cover. This attenuation factor was calibrated on site before the lighting-strategy comparison by simultaneously logging one PPFD sensor outdoors and one PPFD sensor inside the greenhouse for a day and computing the average ratio of indoor to outdoor daily PPFD. The resulting value was then fixed during the validation experiment and used for all treatments and both replicates.

\begin{figure}[!t]
    \centering

    \includegraphics[width=\columnwidth]{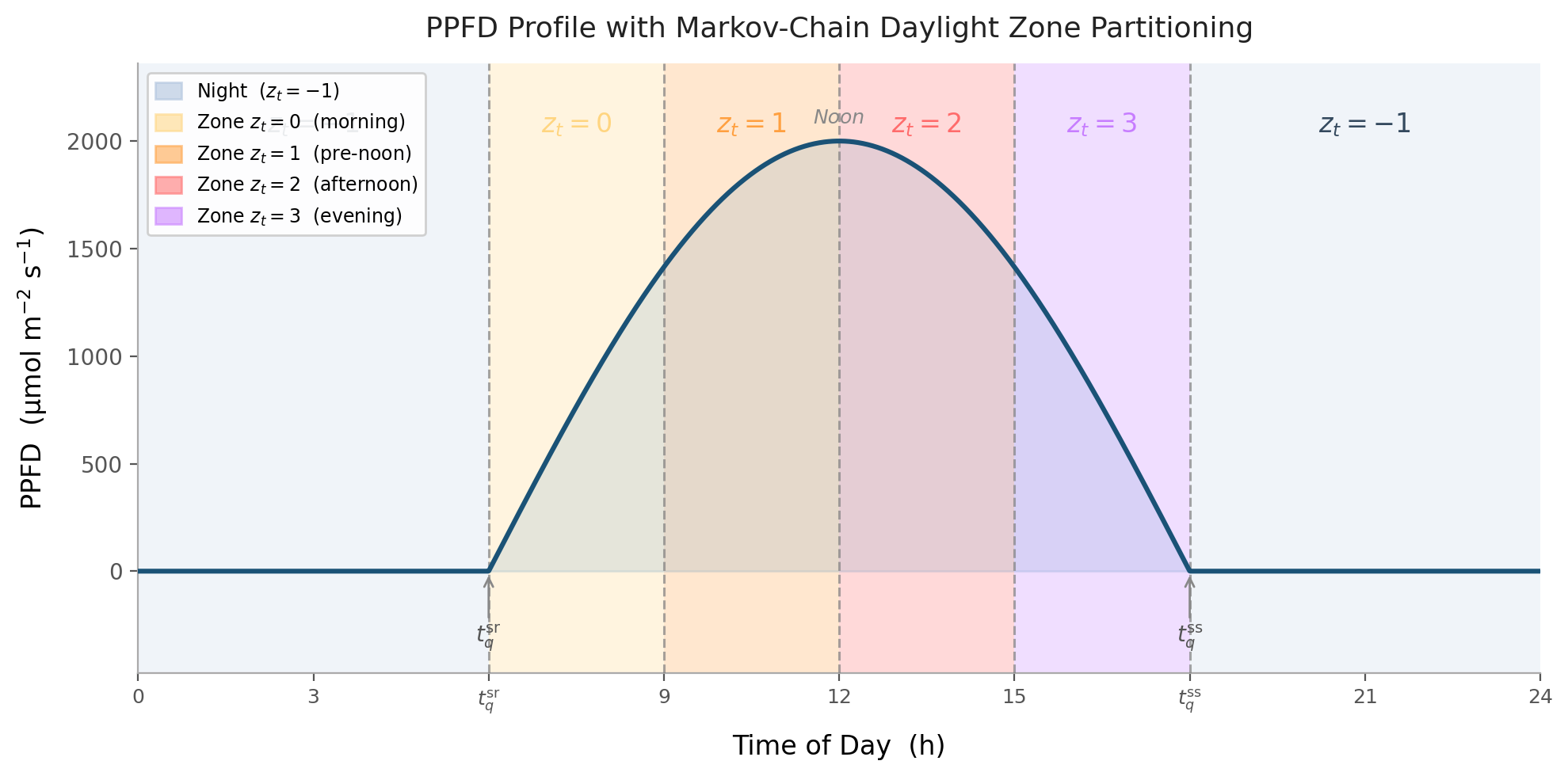}
    \caption{Partitioning of a representative day into Markov-chain contexts. The interval between sunrise $t_q^{\mathrm{sr}}$ and sunset $t_q^{\mathrm{ss}}$ is divided into four equal-duration daylight zones $z_t \in \{0,1,2,3\}$ via~\eqref{eq:zone_fraction}--\eqref{eq:zone_index}, while nighttime samples are assigned $z_t = -1$. A typical sunlight PPFD profile is overlaid to illustrate how the zones capture different phases of the diurnal cycle (morning ramp-up, pre-noon, afternoon, and evening decay).}
    \label{fig:daylight_zones}
\end{figure}

A binary day-night indicator is defined at each 5-minute time step:
\begin{equation}
d_t =
\begin{cases}
1, & \text{if } \theta_t < 90^{\circ} \text{ and } \GHI_t > 0, \\
0, & \text{otherwise},
\end{cases}
\end{equation}
where $\theta_t$ denotes the solar zenith angle. For each day, sunrise and sunset are identified as the first and last daytime samples, respectively. The daylight interval is then partitioned into four equal-duration zones (see Fig. \ref{fig:daylight_zones}). For a sample belonging to day $q$,
\begin{equation}
f_t = \frac{t - t_q^{\mathrm{sr}}}{t_q^{\mathrm{ss}} - t_q^{\mathrm{sr}}},
\label{eq:zone_fraction}
\end{equation}
and the daylight zone is
\begin{equation}
z_t = \min\!\left(3,\;\left\lfloor 4 f_t \right\rfloor\right),
\label{eq:zone_index}
\end{equation}
while nighttime samples are assigned $z_t = -1$.

The context used to condition the Markov transitions is defined by the calendar month, the day--night indicator, and the daylight zone:
\begin{equation}
c_t = \left(m_t,\, d_t,\, z_t\right).
\end{equation}
This yields 60 possible contexts (12 months $\times$ (1 nighttime category + 4 daytime zones)).

Because PPFD is continuous, it is discretized into 60 bins. The bin edges are obtained from the empirical quantiles of the positive PPFD values, while zero irradiance is retained explicitly, so that darkness is represented by the first bin. The discrete state is defined as
\begin{equation}
s_t = \mathcal{B}\!\left(\PPFD_t\right), \qquad s_t \in \{0, 1, \dots, 59\},
\end{equation}
where $\mathcal{B}(\cdot)$ is the binning operator. Each bin is represented by its midpoint, with the first bin anchored at zero.

A first-order Markov model is then estimated in a context-dependent form, where the transition from the current PPFD state to the next PPFD state is conditioned on the context of the \emph{next} time step. The transition counts are
\begin{equation}
N_{ij}^{(c)} = \sum_t \mathbf{1}\!\left\{ s_t = i,\; s_{t+1} = j,\; c_{t+1} = c \right\},
\end{equation}
where $\mathbf{1}\{\cdot\}$ is the indicator function. For daytime contexts, Laplace smoothing with parameter $\alpha_L = 0.2$ is applied, giving the transition probabilities
\begin{equation}
P_{ij}^{(c)} =
\frac{N_{ij}^{(c)} + \alpha_L}
     {\sum_{j'=0}^{59}\!\left(N_{ij'}^{(c)} + \alpha_L\right)}.
\end{equation}
For nighttime, all transitions are forced to the zero-PPFD bin as
\begin{equation}
P_{ij}^{(c)} =
\begin{cases}
1, & \text{if } j = 0, \\
0, & \text{otherwise},
\end{cases}
\quad \text{for nighttime contexts.}
\end{equation}

To support prediction at future timestamps, a day-of-year lookup table of typical sunrise and sunset times is built by averaging the observed sunrise and sunset minutes across the training years. This lookup table is used to determine the day-night indicator and daylight zone of a future 5-minute step.

Let $r_j$ denote the representative PPFD value of bin $j$. Given the current PPFD state and the context of the next time step, the one-step-ahead expected PPFD forecast is
\begin{equation}
\widehat{\PPFD}_{t+1}
\;=\;
\sum_{j=0}^{59} P_{s_t j}^{(c_{t+1})}\, r_j.
\label{eq:ppfd_forecast}
\end{equation}
This formulation yields a nonhomogeneous first-order Markov chain whose transition law varies with season and within-day solar position, thereby preserving the simplicity of Markov dynamics while capturing the strong diurnal and seasonal structure of solar radiation.

\subsection{Shared Notation and Unified Optimization Template}
\label{subsec:opt_notation}

The optimization-based strategies (prediction-based and both peak-shift variants) share the following notation. Over a control horizon of $H$ five-minute steps, the decision variable $x_t \in \mathbb{R}_{\ge 0}$ is the LED PPFD command at step~$t$ ($\mu$mol\,m$^{-2}$\,s$^{-1}$), stacked into the decision vector $\mathbf{x} = [x_1, \dots, x_H]^\top \in \mathbb{R}_{\ge 0}^{H}$. Likewise, $\mathbf{c} = [c_1,\dots,c_H]^\top$, $\mathbf{u}^{\max} = [u_1^{\max},\dots,u_H^{\max}]^\top$, and $\mathbf{s}^{\text{clip}} = [s_1^{\text{clip}},\dots,s_H^{\text{clip}}]^\top$ are the vectors of cost coefficients, per-step LED headroom, and clipped sunlight predictions, respectively. The variable $s_t$ is the predicted indoor sunlight PPFD at step~$t$ obtained from~\eqref{eq:ppfd_forecast}. The remaining model parameters are summarized in Table~\ref{tab:opt_params}.

\begin{table}[!t]
\centering
\caption{Optimization parameters shared by the prediction-based and peak-shift strategies.}
\label{tab:opt_params}
\renewcommand{\arraystretch}{1.1}
\setlength{\tabcolsep}{4pt}
\begin{tabularx}{\linewidth}{l X l}
\toprule
\textbf{Symbol} & \textbf{Meaning}  & \textbf{Value} \\
\midrule
$\Delta t$              & Control step length [s]                                                 & $300$ \\
$\bar{u}$               & Max LED PPFD [$\mu$mol\,m$^{-2}$\,s$^{-1}$]                 & $261.29$ \\
$\bar{P}$               & Assumed max plant-absorbable PPFD [$\mu$mol\,m$^{-2}$\,s$^{-1}$]         & $500$ \\
$s_t$                   & Predicted sunlight PPFD at step $t$                             & forecast \\
$D_0$                   & Cumulative DLI so far [mol\,m$^{-2}$]                              & measured \\
$D_{\mathrm{tar}}$      & Daily DLI target [mol\,m$^{-2}$]                                   & $17$ \\
$\eta$                  & LED photon efficacy [$\mu$mol/J]                              & $2.7$ \\
$\pi_t$                 & Electricity price [\$/kWh]                              & TOU \\
$\lambda$               & Peak-penalty weight                                    & $\{0,\,0.2\}$ \\
\bottomrule
\end{tabularx}
\end{table}

The TOU tariff was \$0.10/kWh from 00:00-06:59, \$0.14/kWh from 07:00-15:59 and 21:00-23:59, and \$0.20/kWh from 16:00-20:59.



Photosynthesis saturates above the canopy absorption limit $\bar{P}$, so photons above it are not expected to contribute to DLI. We therefore clip the sunlight prediction as
\begin{equation}
s_t^{\text{clip}} = \min(s_t,\, \bar{P}),
\label{eq:sun_clip}
\end{equation}
and define the per-step LED headroom
\begin{equation}
u_t^{\max} = \min\!\Big(\bar{u},\; \max\!\big(0,\; \bar{P} - s_t^{\text{clip}}\big)\Big),
\label{eq:led_headroom}
\end{equation}
so LED output cannot exceed what the canopy can still absorb. The LED-supplied DLI still required is
\begin{equation}
\DLI_{\text{req}}
=
\max\!\left(
0,\;
D_{\text{tar}} - D_0 
\right),
\label{eq:dli_req}
\end{equation}
and the cost of one unit of LED PPFD applied for one step is
\begin{equation}
c_t \;=\; \pi_t \cdot \frac{\Delta t}{\eta \cdot 3.6 \times 10^{6}},
\label{eq:cost_coeff}
\end{equation}
since producing $x_t$ requires $x_t/\eta$ W/m$^2$ of electrical power and $1~\mathrm{kWh}=3.6\times 10^{6}$~J. Stacking vectors $\mathbf{c}, \mathbf{u}^{\max}, \mathbf{s}^{\text{clip}}$ componentwise, every optimization-based strategy is an instance of the unified LP
\begin{equation}
\begin{aligned}
\min_{\mathbf{x},\, p} \quad
& \mathbf{c}^{\top}\mathbf{x} \;+\; \lambda \cdot p \\
\text{s.t.}\quad
& \mathbf{0} \le \mathbf{x} \le \mathbf{u}^{\max}, \\
& \tfrac{\Delta t}{10^{6}}\,\mathbf{1}^{\top}(\mathbf{x} + \mathbf{s}^{\text{clip}}) \;\ge\; \DLI_{\text{req}}, \\
& \mathbf{x} \le p\,\mathbf{1}, \qquad p \ge 0,
\end{aligned}
\label{eq:lp_unified}
\end{equation}
parameterized by the peak-penalty weight $\lambda$ and by horizon start. The prediction-based strategy uses $\lambda=0$; the peak-shift variants use $\lambda=0.2$ with different horizon starts.

\subsection{Description of Our Six Control Strategies}\label{sec:strategy}

In this section, we summarize the six lighting control groups we implemented.

\subsubsection{\textbf{Strategy 1: Always-On (Upper-Bound Baseline)}}
\label{subsec:always_on}
The LED array runs at $x_t = \bar{u}$ for $t=1,\dots,H$. Ignoring sunlight and TOU pricing, this strategy upper-bounds energy consumption among active strategies.

\subsubsection{\textbf{Strategy 2: Always-Off (Sunlight-Only Control)}}
\label{subsec:always_off}
No supplemental lighting is provided ($x_t=0$). This treatment isolates the marginal biological contribution of supplemental lighting.

\subsubsection{\textbf{Strategy 3: Prediction-Based}}
\label{subsec:prediction_based}

Setting $\lambda=0$ in~\eqref{eq:lp_unified} and anchoring the horizon at 04:30.

At the optimum, LED output is placed greedily into the cheapest and least-sunlit steps until the DLI constraint is tight. The LP is re-solved every $\Delta t=300$\,s, so forecast errors at one step are absorbed by subsequent re-planning.

\subsubsection{\textbf{Strategy 4: Heuristic (Forecast-Free)}}
\label{subsec:heuristic}
The heuristic controller tracks the DLI target by setting the LED PPFD to the uniform catch-up rate required over the remaining steps. At step $t$, let $D_0^t$ denote the total accumulated DLI (sunlight plus prior LED contribution) measured up to that point. The LED command is then
\begin{multline}
    x_t \;=\; \min\!\left(\bar{u},\;\max\!\left(0,\;
    \frac{10^6\,\bigl(D_{\mathrm{tar}} - D_0^t\bigr)}{\Delta t\,(H - t)}
    \right)\right), \\
    t = 1, \dots, H-1,
    \label{eq:heuristic}
\end{multline}

\subsubsection{\textbf{Strategies 5 and 6: Peak-Shift Variants}}
\label{subsec:peak}
Both peak-shift variants instantiate~\eqref{eq:lp_unified} with
$\lambda = 0.2$, so the auxiliary $p \ge 0$, acting as the
$\ell_\infty$-norm of $\mathbf{x}$, penalizes the tallest LED spike and
flattens the schedule (useful when demand charges apply). The two
variants differ in whether TOU price information enters the objective.
\textbf{Strategy~5 (cost-aware peak-shift)} keeps the full cost
coefficients $c_t$ from~\eqref{eq:cost_coeff}, so the objective
$\mathbf{c}^{\top}\mathbf{x} + \lambda p$ trades off TOU energy cost
against peak flattening. \textbf{Strategy~6 (peak-only)} sets
$c_t = 0$ for all $t$, collapsing the objective to $\lambda . p$ alone;
the optimizer then minimizes peak LED power subject only to the DLI
constraint, with no preference for off-peak windows. The contrast
isolates the contribution of the TOU-aware cost term once peak
flattening is already being enforced.

\section{Results}
\label{sec:results}
Each strategy was replicated twice in diagonally opposite zones (Fig.~\ref{fig:greenhouse_layout}). Replicate~1 ran 20~February--19~March~2026 and Replicate~2 ran 20~March--15~April~2026. Each treatment used 24 plants per cultivar across three cultivars (Hydrolique, Rabello, Sungari), for 72~plants per treatment. We organize the analysis around three questions: (i) do the strategies deliver the DLI target and how do they shape the LED schedule; (ii) what does each cost; and (iii) how do those costs translate into biological outcome.

\subsection{DLI Compliance and LED Scheduling Behavior}
\label{subsec:timing}

\begin{figure}[htbp]
    \centering
    \includegraphics[width=\columnwidth]{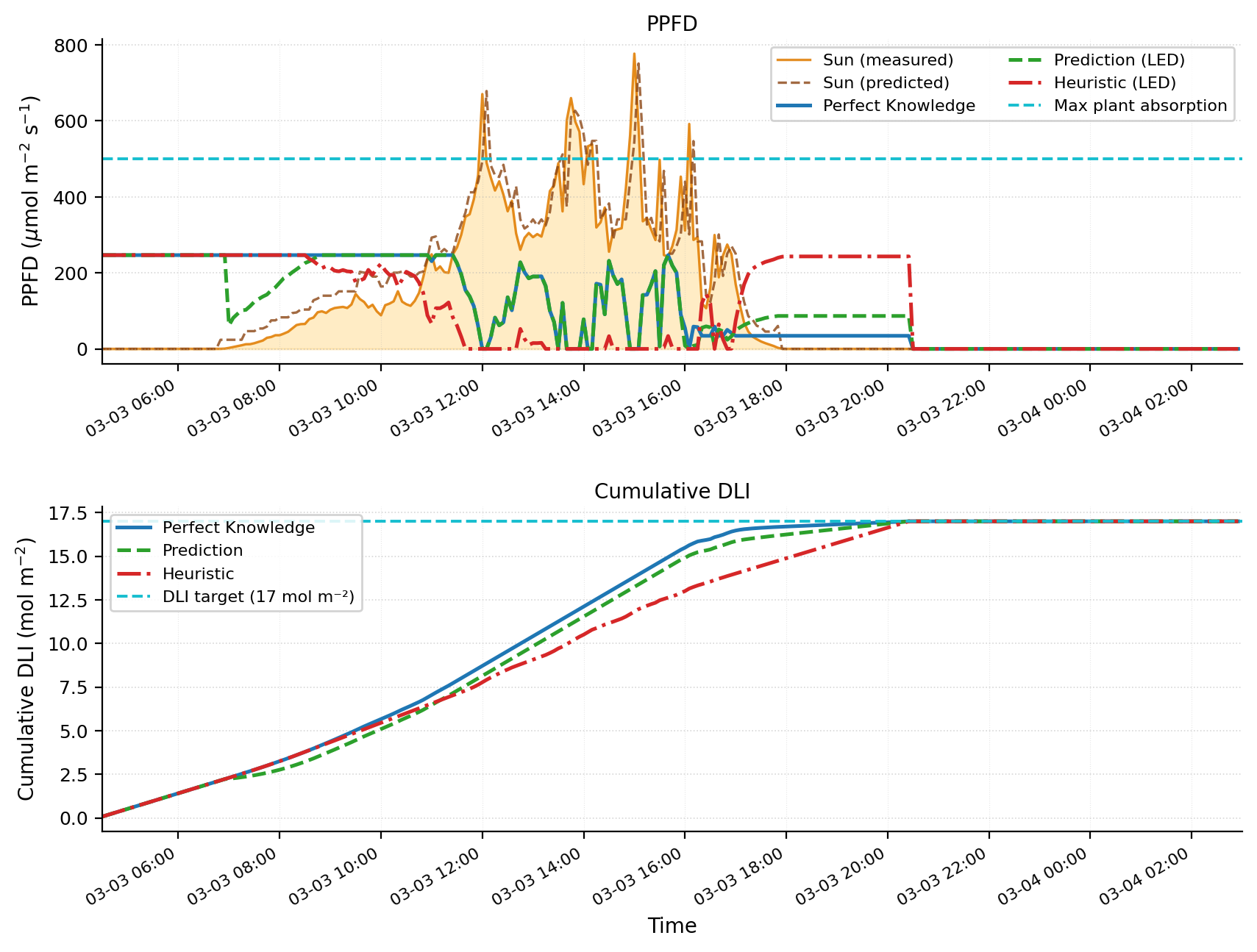}
    \caption{LED commands and cumulative DLI for the adaptive strategies on 3~March~2026. The x-axis shows local time in MM-DD HH:MM format. The prediction-based and heuristic controllers both converge to the 17~mol\,m$^{-2}$ target. \emph{Perfect knowledge} solves the same LP as \emph{prediction-based} but replaces the Markov-chain forecast with the sunlight actually observed that day, so the two would overlap under a perfect forecast; the gap between them quantifies the cost of forecast error.}
    \label{fig:smart_strategies}
\end{figure}

\begin{figure}[htbp]
    \centering
    \includegraphics[width=\columnwidth]{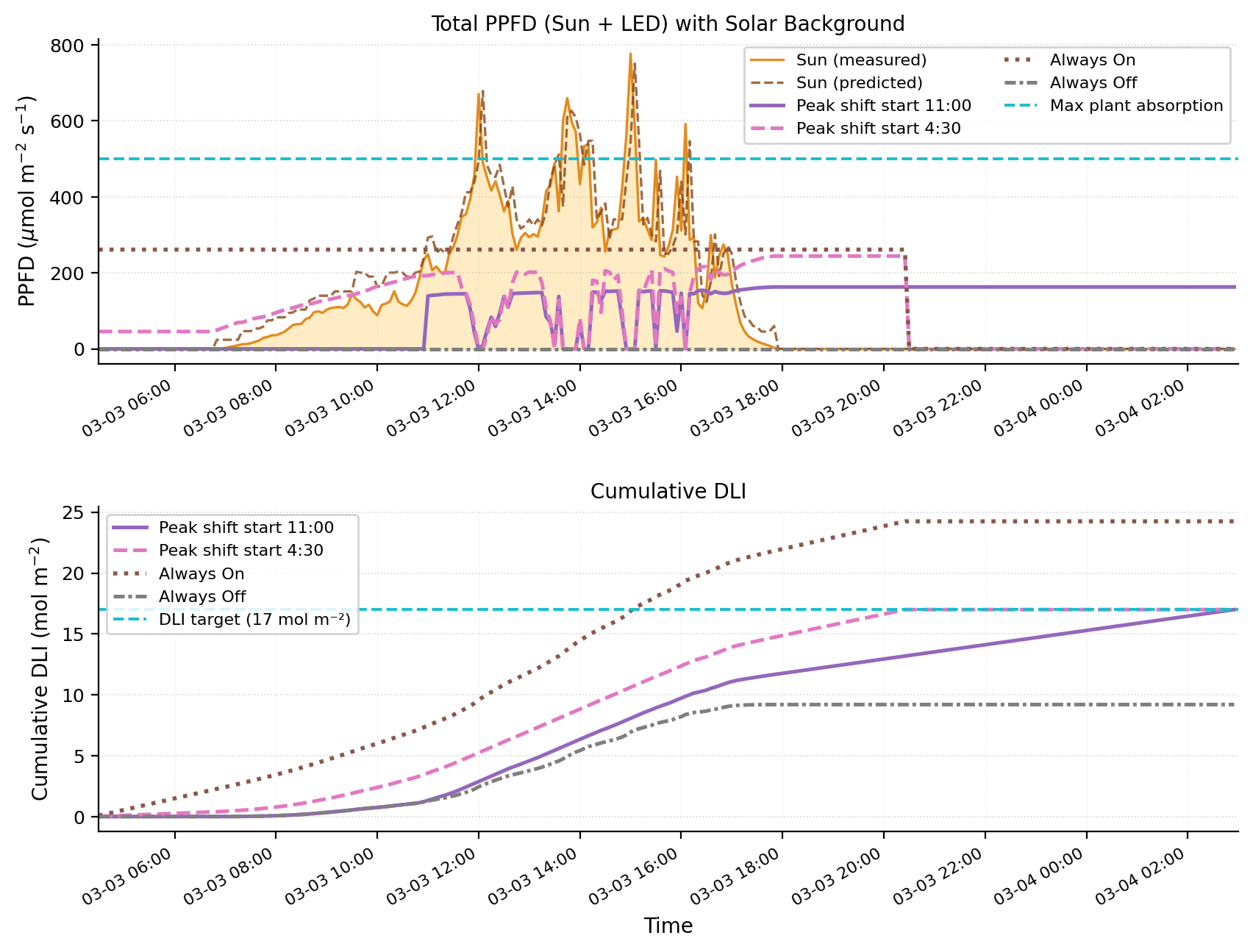}
    \caption{LED commands and cumulative DLI for baseline and peak-shift strategies on 3~March~2026. The x-axis shows local time in MM-DD HH:MM format. Always-on overshoots the target, always-off remains below it, and the peak-shift variants converge at different rates depending on horizon start.}
    \label{fig:benchmark_strategies}
\end{figure}

Figs.~\ref{fig:smart_strategies} and~\ref{fig:benchmark_strategies} show total canopy PPFD and cumulative DLI on a representative day (3~March~2026). \emph{Always-on} holds LED PPFD at $\bar{u}$ independent of ambient conditions, overshooting the 17~mol\,m$^{-2}$ target well before the end of the photoperiod. \emph{Always-off} never reaches the target, confirming that sunlight alone is insufficient at this site and season. The \emph{prediction-based} and \emph{heuristic} strategies both converge to the target (Fig.~\ref{fig:smart_strategies}) but via different paths: the prediction-based controller anticipates forecasted afternoon sunlight, suppresses LEDs during it, and concentrates supplemental light in the cheapest and darkest morning/evening windows, while the heuristic distributes LED output more uniformly and spends a larger fraction of its energy during hours when solar irradiance would have been adequate. The two \emph{peak-shift} variants differ primarily in horizon placement: the early-start variant front-loads supplemental light into the pre-noon window, while the midday variant packs its LED energy into a compressed horizon centered on solar noon, wasting more headroom to canopy saturation and overlapping more with shoulder pricing.

\subsection{Energy--Yield Tradeoff}
\label{subsec:energy_yield_tradeoff}

Table~\ref{tab:yield_per_dollar} summarizes the joint energy-cost and crop-yield outcomes across all six strategies. Daily energy cost ranged from \$0 for always-off to \$0.2332/m$^2$ for always-on. Among the active strategies, the prediction-based controller reduced cost to \$0.0807/m$^2$, corresponding to a 65\% reduction relative to always-on, while producing the highest dry mass among the adaptive strategies, 1.531~g plant$^{-1}$. The heuristic controller achieved a similar cost level, with only a 6\% premium over prediction-based control (\$0.0858/m$^2$), but its dry mass was lower at 1.302~g plant$^{-1}$. The two peak-shift variants achieved the lowest active-strategy costs, \$0.0715--\$0.0738/m$^2$, but also produced lower dry mass, 1.252--1.264~g plant$^{-1}$. Always-on achieved the highest absolute yield, 2.190~g plant$^{-1}$, because it delivered the most supplemental light, but it did so at the highest cost. The canopy-usefulness limit $\bar{P}$ in~\eqref{eq:sun_clip}--\eqref{eq:led_headroom} is an instantaneous PPFD clipping limit used by the optimized experiments (strategy 3,4,5, and 6 (see section \ref{sec:strategy}), Always on and always off have no PPFD clipping limit. Strategies can therefore produce different dry masses even when they avoid crediting photons above $\bar{P}$, because they differ in total delivered DLI, the timing of supplemental light relative to natural sunlight, and the accumulated plant response over the full growth period. In particular, always-on delivers substantially more total light than the adaptive strategies, which explains its higher dry mass, while the prediction-based strategy provides the highest dry mass among the adaptive strategies at much lower cost.

\begin{table}[!t]
    \centering
    \small
    \setlength{\tabcolsep}{4pt}
    \caption{Yield-per-dollar efficiency across lighting strategies (24 plants $\times$ 3 cultivars = 72 plants per treatment). Rows ordered by ascending $\eta_\$$.}
    \label{tab:yield_per_dollar}
    \renewcommand{\arraystretch}{1.1}
    \begin{tabular}{lccc}
        \toprule
        \textbf{Strategy} & \textbf{Dry Mass (g)} & \textbf{Cost (\$/m\textsuperscript{2})} & $\boldsymbol{\eta_\$}$ \textbf{(g/\$)} \\
        \midrule
        Always-off                   & 0.448 & 0           & ---         \\
        Always-on                    & 2.190 & 0.2332      & 9.391       \\
        Heuristic                    & 1.302 & 0.0858      & 15.170  \\
        Peak 11:00 ($\lambda=0.2$)   & 1.264 & 0.0738      & 17.127   \\
        Peak 04:30 ($\lambda=0.2$)   & 1.252 & 0.0715      & 17.520   \\
        Prediction-based             & 1.531 & 0.0807      & 18.976   \\
        \bottomrule
    \end{tabular}
\end{table}

Cost and yield alone therefore lead to different rankings: always-on maximizes yield, whereas peak-shift minimizes active-strategy cost. To compare both objectives jointly, we use the yield-per-dollar efficiency $\eta_\$ = Y/C$, where $Y$ is dry mass per treatment area and $C$ is daily energy cost per unit area. The prediction-based strategy achieved the highest efficiency, 18.976~g/\$, more than double the always-on value of 9.391~g/\$. The peak-shift variants ranked next, with 17.127--17.520~g/\$, mainly by reducing cost rather than increasing yield. The heuristic strategy achieved 15.170~g/\$, reflecting its slightly higher cost and lower dry mass relative to prediction-based control. Overall, the prediction-based controller provides the strongest cost-yield tradeoff, while peak-shift may be preferable when demand reduction is prioritized and always-on may be preferable when maximizing absolute yield outweighs energy cost.

\section{Conclusion}
\label{sec:conclusion}
 
This paper presented a controlled, side-by-side comparison of six LED lighting control strategies deployed simultaneously in a single greenhouse at the Hampton Roads AREC of Virginia Tech. 
Three findings stood out. First, the prediction-based strategy, which solves an energy-minimizing linear program at each $\Delta t = 300$\,s re-planning step using a context-conditioned Markov-chain solar forecast, achieved the DLI target of 17~mol\,m$^{-2}$ at a daily cost of \$0.0807/m$^2$, which is a 65\% reduction relative to the always-on baseline (\$0.2332/m$^2$) while delivering 1.531~g of dry mass per plant, the highest among the adaptive strategies. Second, the heuristic strategy, which requires no forecasting infrastructure, incurred only a modest 6\% cost premium over the prediction-based approach (\$0.0858 vs.\ \$0.0807/m$^2$), making it an attractive option for operators who prioritize deployment simplicity over marginal cost savings. Third, the peak-shift variants achieved the lowest absolute energy cost among active strategies (\$0.0715--\$0.0738/m$^2$) by redistributing LED demand into off-peak tariff windows, though at the cost of a 17--18\% reduction in dry mass relative to the prediction-based strategy. To integrate these trade-offs into a single operational metric, we introduced the yield-per-dollar efficiency $\eta_\$$, which ranked the prediction-based strategy highest (18.976~g/\$), more than double the always-on value (9.391~g/\$), demonstrating that forecast-driven control can simultaneously reduce energy expenditure and improve economic return.
 


\bibliography{ifacconf}

\end{document}